\documentclass[
    ,final 
  ]
  {aipproc}

\usepackage{amssymb}
\layoutstyle{6x9}

\begin{document}

\title{The Swift SFXT monitoring campaign:
the IGR~J16479-4514 outburst in 2009}


\author{V. La Parola}{
  address={INAF-IASF Via U.\ La Malfa 153, I-90146 Palermo, Italy}
}

\author{L. Ducci}{
  address={INAF-IASF Via E.\ Bassini 15, I-20133 Milano, Italy}
  ,altaddress={Universit\`a degli Studi dell'Insubria, Via Valleggio 11, I-22100 Como, Italy}
}

\author{P. Romano}{
  address={INAF-IASF Via U.\ La Malfa 153, I-90146 Palermo, Italy}
}

\author{L. Sidoli}{
  address={INAF-IASF Via E.\ Bassini 15, I-20133 Milano, Italy}
}

\author{G. Cusumano}{
  address={INAF-IASF Via U.\ La Malfa 153, I-90146 Palermo, Italy}
}

\author{S. Vercellone}{
  address={INAF-IASF Via U.\ La Malfa 153, I-90146 Palermo, Italy}
}

\author{V. Mangano}{
  address={INAF-IASF Via U.\ La Malfa 153, I-90146 Palermo, Italy}
}

\author{J. A. Kennea}{
  address={Department of Astronomy and Astrophysics, Pennsylvania State 
             University, University Park, PA 16802, USA}
}

\author{H. A. Krimm}{
  address={NASA/Goddard Space Flight Center, Greenbelt, MD 20771, USA}
  ,altaddress={Universities Space Research Association, Columbia, MD, USA}
}

\author{D. N. Burrows}{
  address={Department of Astronomy and Astrophysics, Pennsylvania State 
             University, University Park, PA 16802, USA}
}

\author{N. Gehrels}{
  address={NASA/Goddard Space Flight Center, Greenbelt, MD 20771, USA}
}
\begin{abstract}

IGR~J16479-4514 is a member of the Supergiant Fast X-ray transient (SFXT) 
class. We present the light curves of its latest outburst, which occurred on 
January 29, 2009. During this outburst, IGR~J16479-4514 showed prolonged 
activity lasting  several days. The presence of eclipses was successfully
tested.

\end{abstract}

\maketitle


\section{The outburst on 2009 February}
The {\it Swift}/BAT caught a new outburst from the Supergiant Fast X-ray Transient 
IGR J16479-4514 on January 29 \cite{atel1920}. 
{\it Swift} slewed to the target so that the XRT started observing the field at 
06:46:46.9 UT, 819.3 s after the BAT trigger. 
In the days following the outburst, the source was regularly monitored with 
{\it Swift}/XRT, and showed renewed activity on 2009 February 8 \cite{atel1929}.
The XRT data were sought for 
the presence of eclipses, as suggested by \cite{bozzo08}. 
Using the ephemeris from \cite{jain09}, we selected the events inside and 
outside the eclipses, where by `inside the eclipse' we consider the time 
interval between the start of the eclipse as defined by \cite{bozzo08} 
and 0.6 d later. We obtain a net count rate of 
$(6\pm3)\times10^{-3}$~counts s$^{-1}$ (inside) and $0.203\pm0.003$~counts s$^{-1}$ 
(outside). This indicates that the source is in two 
distinct flux levels inside and outside the predicted times of the eclipses 
at the $\sim50\sigma$ level. 
Consistent results are obtained examining the data of the whole 2008 XRT
campaign. We can thus conclude that the XRT data are consistent with the 
presence of an eclipse on the longest baseline so far examined. 

\section{Discussion}

The 3.32 days periodicity \cite{jain09}, if interpreted as the orbital period 
of the binary system, is very difficult to reconcile with the 
mechanisms proposed to explain the SFXTs phenomenon. We can compare the 
out-of-eclipse average X-ray luminosity (L$_{\rm obs}\lesssim10^{35}$ 
erg s$^{-1}$) with the X-ray emission expected from Bondi-Hoyle accretion onto
a neutron star. Assuming a circular orbit and a set of realistic physical 
parameters for the binary system \cite{romano09}, the expected X-ray luminosity 
is $\sim10^{37}$ erg s$^{-1}$, about 2--3 order of magnitude higher than the observed
value, that, on 
the other hand, can be obtained only at a wind 
mass loss rate of \.M$\lesssim10^{-9}M_{\odot}$yr$^{-1}$ (which is not 
reasonable for a 
O8.5 supergiant). A viable explanation to this inconsistency could be that the 
3.32 days periodicity is not orbital, but it is the 
time interval between the periodically recurrent flares when the neutron star 
passes through the preferential plane for the outflowing wind from the 
supergiant; thus, the true orbital period can be much longer 
than this periodicity \cite{sidoli07}.


\begin{figure}
  \includegraphics[width=6.3cm,angle=0]{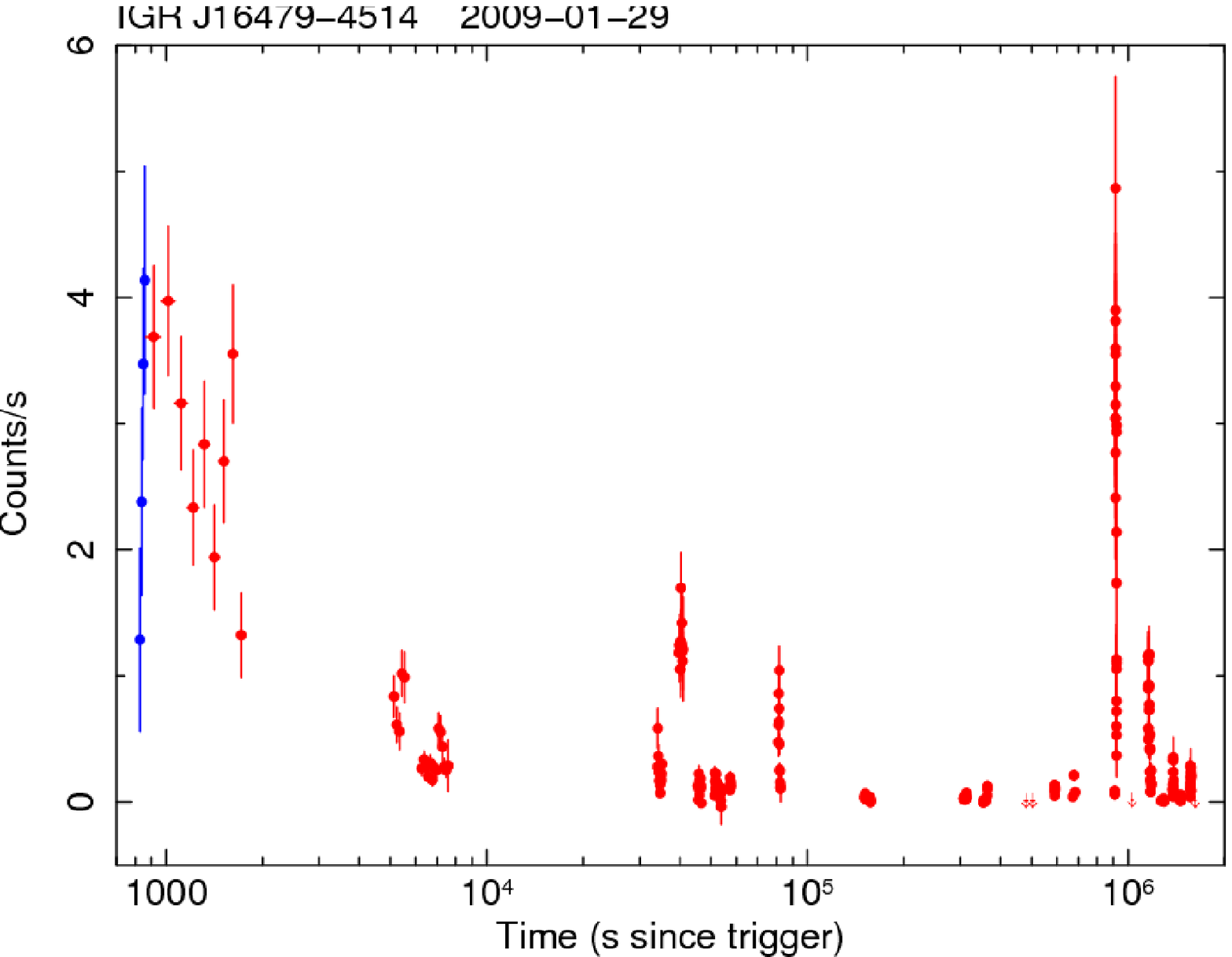}
  \includegraphics[width=7cm,angle=0]{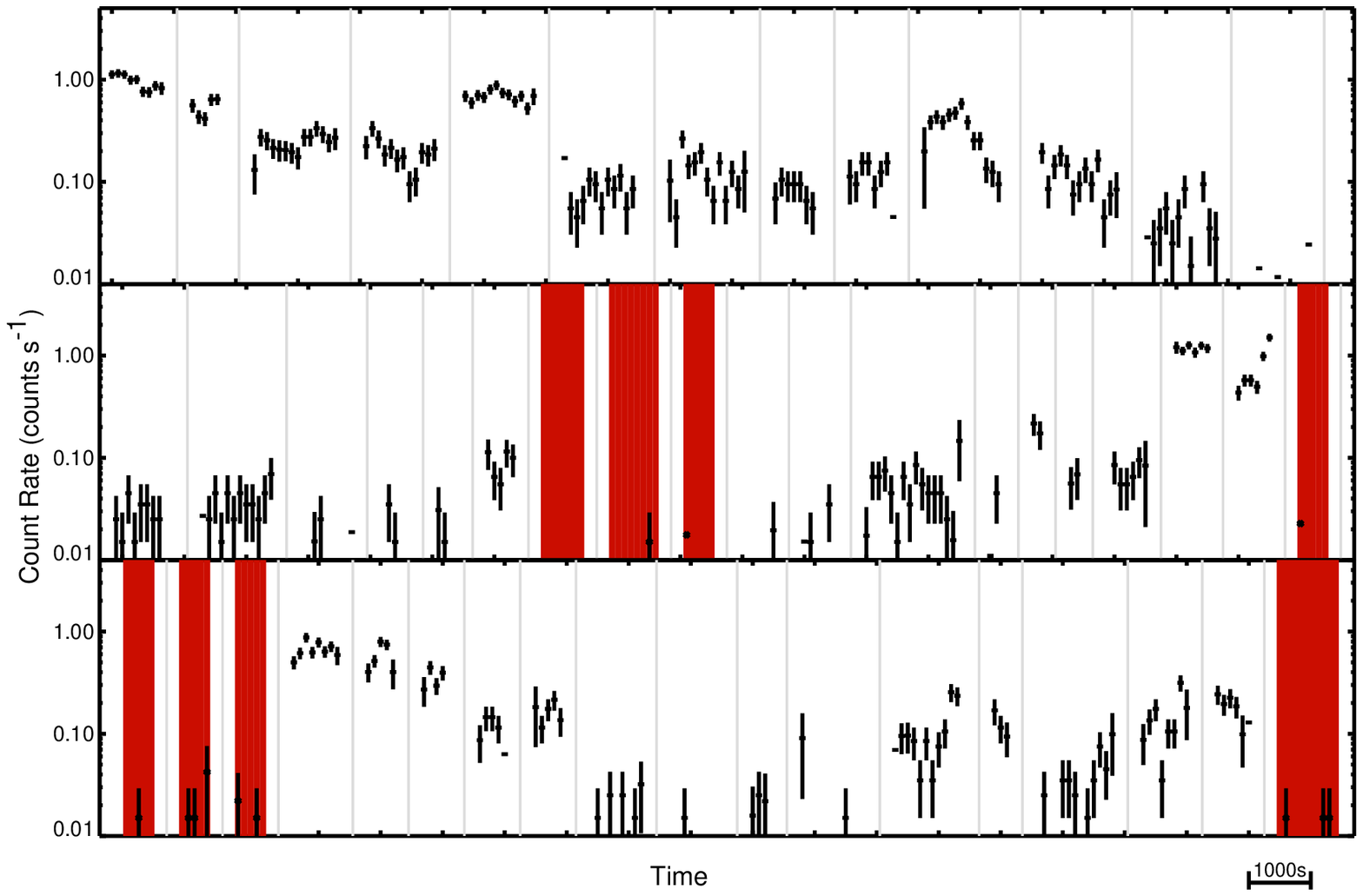}
  \caption{{\bf Left Panel}: {\it Swift}-XRT (0.2-10 keV) lightcurve of the January 29
  outburst of IGR J16479-4514. 
  {\bf Right Panel}: Montage of the outburst time sequences. 
  Non-observing intervals and orbital gaps have 
  been cut out from the time axis and replaced by thick grey vertical bars. 
  The red strips mark the position of the predicted eclipse times. Each point 
  represents a 100 s time bin. As a reference, on the bottom-right we show 
  the 1000s time unit.}
\end{figure}



\begin{theacknowledgments}
This work was supported by contract ASI/INAF I/088/06/0 and I/023/05/0 in Italy, by
NASA contract NAS5-00136 at PSU.
\end{theacknowledgments}



\bibliographystyle{aipproc} 


\begin{thebibliography}{9}

\bibitem{atel1920}
Romano, P.; Sidoli, L.; Mangano, V.; et al. \emph{ATel 1920}, 2009

\bibitem{atel1929}
La Parola, V.; Romano, P.; Sidoli, L.; et al. \emph{ATel 1929}, 2009

\bibitem{bozzo08}
Bozzo, E.; Stella, L.; Israel, G.; Falanga, M.; Campana, S. \emph{MNRAS}, 
\textbf{391}, 108, 2008

\bibitem{jain09}
Jain, C.; Paul, B.; Dutta, A. \emph{MNRAS} 
\textbf{397}, 11, 2009

\bibitem{romano09}
Romano, P.; Sidoli, L.; Cusumano, G.; La Parola, V.; Vercellone, S.; Pagani, C.;
 Ducci, L.; Mangano, V.; Cummings, J.; Krimm, H. A.; Guidorzi, C.; Kennea, J. A.; 
 Hoversten, E. A.; Burrows, D. N.; Gehrels, N. \emph{MNRAS} 
\textbf{399}, 2021, 2009

\bibitem{sidoli07}
Sidoli, L.; Romano, P.; Mereghetti, S.; Paizis, A.; Vercellone, S.; 
Mangano, V.; Gotz, D. \emph{A\&A} 
\textbf{476}, 1307, 2007

\end{thebibliography}


\end{document}